\documentclass[dvips]{article}
\newcommand{\hess}{\textit{H.E.S.S.}}
\newcommand{\fermi}{\textit{Fermi-LAT}}
\newcommand{\dixdix}{\textit{1RXS~J101015.9-311909}}
\newcommand{\trezedouze}{\textit{1ES~1312-423}}
\newcommand{\shbl}{\textit{SHBL~J001355.9-185406}}
\newcommand{\hesscandidate}{\textit{HESS~J1943+213}}
\newcommand{\aplib}{\textit{AP~Lib}}
\usepackage{icrc2011}

\title{Recent \hess\ results on extra-galactic sources}

\shorttitle{M.Cerruti - Recent \hess\ results on extra-galactic sources}

\authors{Matteo Cerruti$^{1}$, \it{on behalf of the \hess\ collaboration}}
\afiliations{$^1$LUTh, Observatoire de Paris, CNRS, Universit$\acute{e}$ Paris Diderot;\\ 5 Place Jules Janssen, 92190, Meudon, France} 
\email{matteo.cerruti@obspm.fr}

\abstract{Since the beginning of scientific operations in 2003, more than 20
extra-galactic sources have been detected with \hess\ (High Energy Stereoscopic System) at very high energies (VHE): apart from the starburst galaxy \textit{NGC~253}, all of them are
active galactic nuclei (AGNs). BL Lac objects are by far the dominant AGN sub-class
in the \hess\ extra-galactic sky, but two radio-galaxies and one flat-spectrum radio
quasar (FSRQ) have been detected as well. The study of extra-galactic VHE emitters will be
improved in the near future by the completion of a fifth 24m-diameter telescope
(\hess\ \textit{II}).
In this talk, a review of the AGNs seen by H.E.S.S. will be given, including the
latest results achieved, namely the detection of VHE emission from the blazars
\dixdix, \shbl, \trezedouze, \aplib\ and the BL Lac candidate \hesscandidate.}

\begin{document}
\maketitle

\section{The \hess\ extra-galactic sky}

\textit{H.E.S.S.}, a system of four atmospheric Cherenkov telescopes sensitive to $\gamma$-ray photons above $100$ GeV, has detected more than 20 extra-galactic sources since the beginning of scientific operations in 2003. \\ Apart from the starburst galaxy \textit{NGC~253}, the VHE extra-galactic sky seen by \hess\ is entirely composed of AGNs. Among these, blazars (BL Lac objects and FSRQ) are by far the most important sub-class. In the unified AGN model \cite{Urry}, these objects are radio-loud AGNs whose relativistic jet is pointed towards the observer. The emitting region is thought to be a very dense zone inside the jet, filled with a high-energy particle ($e^\pm$) population and a magnetic field of the order of $10^{-2}$-$10^{-1}$G. The high energy emission is usually interpreted as inverse Compton scattering of synchrotron photons (synchrotron self-Compton model, SSC), or external photons (external inverse Compton model), off the high energy particles in the emitting region. The measured flux is enhanced by relativistic effects.\\
Apart from the FSRQ \textit{PKS~1510-089} \cite{1510} and the two radio-galaxies \textit{Cen~A} \cite{CenA} and \textit{M~87} \cite{M87,M87bis}, all the AGNs detected with \hess\ are BL Lac objects.\\

\begin{figure}[!t]
  \vspace{5mm}
  \centering
  \includegraphics[width=3.3in]{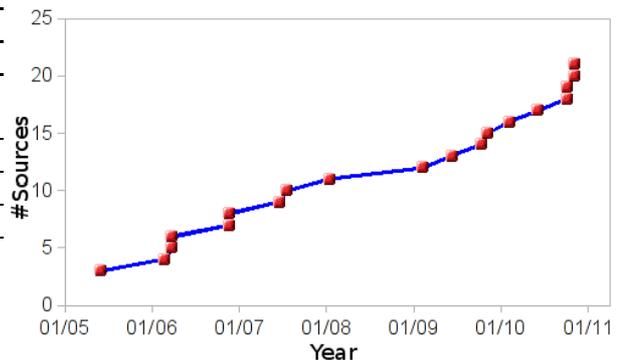}
  \caption{Cumulative number of extra-galactic sources detected by \hess\ as a function of the year}
  \label{simp_fig}
 \end{figure}

The list of the extra-galactic sources detected by \hess\ up to now is reported in Table 1, while the integrated number of extra-galactic sources as a function of the time is shown in Fig. \ref{simp_fig}.
Recently, an important improvement has been achieved through the developement of enhanced analysis techniques \cite{Yvonne,Mathieu,Ohm}, which permit the detection of faint $\gamma$-ray sources with shorter observation time compared to the standard Hillas reconstruction technique \cite{Hillas}.\\
In the following, we concentrate on some of the most recent \hess\ results on extra-galactic sources, namely the detection of VHE emission from the BL Lac objects \textit{SHBL~J001355.9-185406}, \textit{1RXS~J101015.9-311909}, \trezedouze, \aplib\ and the BL Lac candidate \hesscandidate.\\

\begin{table}[t]
\begin{center}
\begin{tabular}{c|c|c}
\hline
& Redshift & Ref. \\
\hline
\textbf{Starburst Galaxies}& &\\
\hline
NGC 253 & $8.14\times10^{-4}$ & \cite{NGC253}\\
\hline
\textbf{Radio Galaxies}& &\\
\hline
M 87 & $0.0044$& \cite{M87,M87bis}\\
Centaurus A & $0.00183$ & \cite{CenA}\\
\hline
\textbf{FSRQs}& &\\
\hline
PKS 1510-089 & $0.36$ & \cite{1510}\\
\hline
\textbf{BL Lacs}& &\\
\hline
SHBL J001355.9-185406 & $0.095$ & \cite{0013}\\
RGB J0152+017 & $0.08$ &\cite{0152}\\
1ES 0229+200& $0.14$ &\cite{0229}\\
1ES 0347-121& $0.188$ &\cite{0347}\\
1ES 0414+009& $0.287$ & \cite{0414}\\
PKS 0447-439& $>0.176$ & \cite{0447}\\
PKS 0548-322& $0.069$ &\cite{0548}\\
1RXS J101015.9-311909& $0.143$ & \cite{Texas}\\
1ES 1101-232& $0.186$ &\cite{1101}\\
Mrk 421& $0.031$ &\cite{Mrk421}\\
1ES 1312-423& $0.105$ & \cite{Texas}\\
AP Lib & $0.049$ & \cite{APLib, APLibtexas}\\
PG 1553+113& $0.43-0.58$ \cite{1553redshift} & \cite{1553,1553bis}\\
HESS J1943+213& $>0.14$ &\cite{1943}\\
PKS 2005-489& $0.071$ &\cite{2005,2005bis}\\
PKS 2155-304& $0.116$ & \cite{2155,2155bis,2155ter}\\
& & \cite{2155quater,2155quinquies,2155sexies}\\
H 2356-309& $0.165$ &\cite{2356,2356bis}\\
\hline 
\end{tabular}
\caption{Extra-galactic sources detected by \hess\ }
\label{tablepippo}
\end{center}
\end{table}

\section{Recent \hess\ results on extra-galactic sources}

\subsection{\shbl\ }

\shbl\ is a high-energy-peaked BL Lac object (HBL) \cite{shblcatalog} located at a redshift of $z=0.095$. The VHE emission from this very weak source is detected by \hess\ at $\sim5$ standard deviations ($\sigma$) above $300$ GeV, during $40$ hours of observation live-time taken between 2008 and 2010 \cite{0013}. The excess map is plotted in Fig. \ref{dixdixplot}. The TeV flux from the source corresponds to $\sim1\%$ of the flux from the Crab nebula (''Crab units''). Following the \hess\ announcement, the presence of a GeV counterpart was claimed by the \fermi\ collaboration \cite{0013Fermi} : the flux of the source above $100$ MeV is $(9\pm7)\times10^{-10}\  \textrm{ph}/\textrm{cm}^2/\textrm{s}$, with a photon index of $1.5\pm0.2$. However, the source is not present in the \fermi\ 2-year catalog (2FGL) \cite{2FGL}.

\subsection{\dixdix\ }
\begin{figure*}[t!]
  \vspace{5mm}
  \centering
  \includegraphics[width=6in]{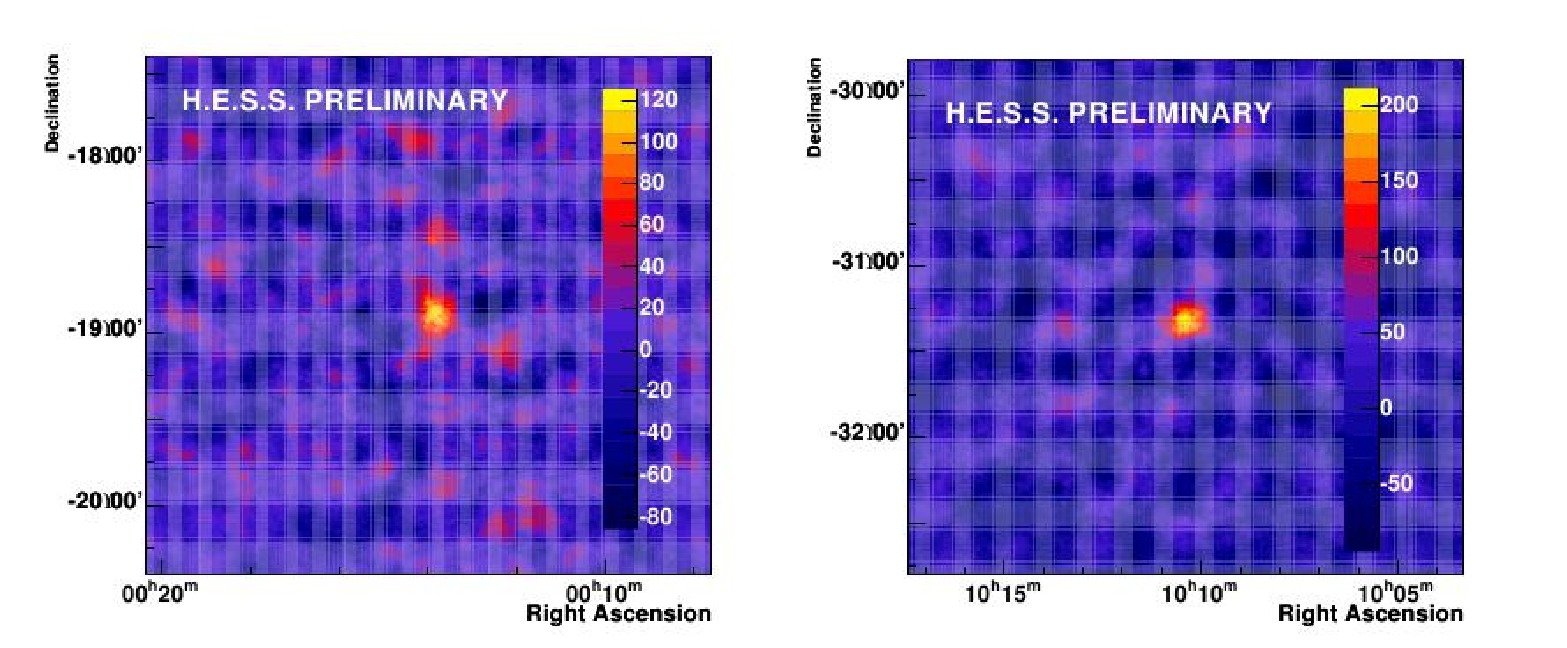}
  \caption{Excess map centered on \shbl\ (left) and \dixdix\ (right). Taken from \cite{Texas}.}
  \label{dixdixplot}
 \end{figure*}

\dixdix\ is a bright X-ray source detected for the first time by \textit{ROSAT} \cite{Rosat}, and identified as a blazar at redshift $z=0.143$ \cite{dizdizid}. The VHE $\gamma$-ray emission from this object has been detected by \hess\ at $7\sigma$, in $49$ hours of observation live-time taken between 2006 and 2010 \cite{Texas}. The excess map is plotted in Fig. \ref{dixdixplot}. The time-averaged photon index of the source is $3.1\pm0.5_{stat}$, with a flux equal to $\sim0.008$ Crab units. The source is present in the 2FGL catalog : the flux between $300$ MeV and $100$ GeV is $(3.5\pm0.9)\times10^{-9}\  \textrm{ph}/\textrm{cm}^2/\textrm{s}$, the photon index being $2.2\pm0.1$. No significant flux variability has been detected in the \hess\ data. 

\subsection{\trezedouze\ }
\begin{figure}[!t]
  \vspace{5mm}
  \centering
  \includegraphics[width=3.5in]{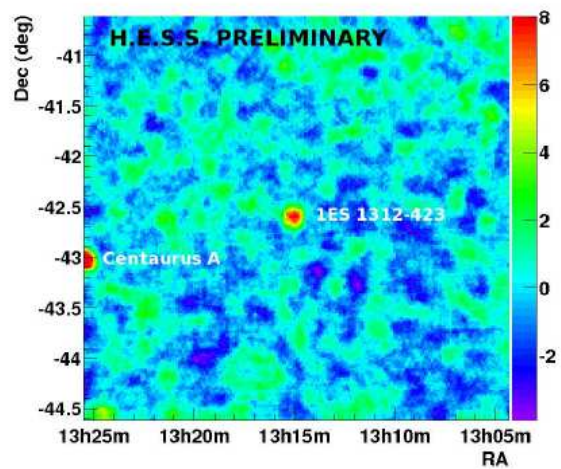}
  \caption{\hess\ significance map of the sky region around \trezedouze\, with the radio-galaxy \textit{Cen A} in the same field-of-view \cite{Texas}}
  \label{trezedouzeplot}
 \end{figure}

\trezedouze\ is an X-ray source classified as a blazar at a redshift of $z=0.105$ \cite{Einstein}. The object was discovered serendipitously in the field of view of \hess\ observations centered on the radio galaxy \textit{Centaurus A} (see Fig.\ref{trezedouzeplot}). The total observation live-time (corrected for the lower acceptance due to the $2^\circ$ offset from the center of the field of view) is $\sim65$ hours. The TeV emission from \trezedouze\ is detected by \hess\ at $\sim7\sigma$ with a flux equal to $0.004$ Crab units \cite{Texas}. The source is not present in the \fermi\ 2-year catalog and no significant variability has been detected in the \hess\ data set.

\subsection{\aplib\ }

\begin{figure*}[t!]
  \vspace{5mm}
  \centering
  \includegraphics[width=5in]{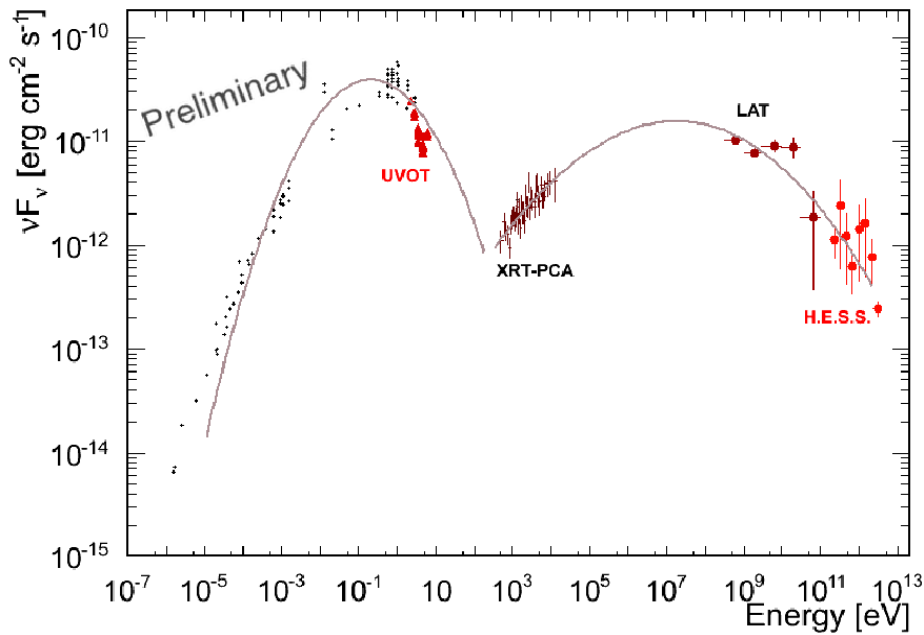}
  \caption{Spectral energy distribution of \aplib. See \cite{APLibtexas} for details.}
  \label{aplibsed}
 \end{figure*}

\aplib\ is a nearby AGN ($z=0.049$), classified as a low-energy-peaked BL Lac object (LBL) \cite{APLibid}. The TeV emission from \aplib\ is detected by \hess\ at $7\sigma$ in $11$ hours of observation live-time taken in 2010 \cite{APLib, APLibtexas}. The VHE spectral index of the source is $\Gamma$=$2.5\pm0.2$, and its flux is $\sim0.02$ Crab units. The object is detected by \fermi\ as well, with an integrated $0.3$-$300$ GeV flux of $(1.9\pm0.1)\times10^{-8} \textrm{ph}/\textrm{cm}^2/\textrm{s}$, and a spectral index of $2.1\pm0.1$ \cite{APLibtexas}. The SED of \aplib\ (see Fig. \ref{aplibsed}\cite{APLibtexas}) shows a very broad high energy bump, which covers the X-ray to VHE $\gamma$-ray energy band. Such a broad component is usually explained by assuming inverse Compton emission on the external photon field (i.e. broad line region, accretion disk), which is thought to be important in LBLs \cite{Ghisellini}. \aplib\ is the third LBL ever detected at VHE (together with \textit{BL~Lac} and \textit{S5~0716+714}).

\subsection{\hesscandidate\ }

\hesscandidate\ is a point-like source detected by \hess\ during the VHE galactic survey \cite{1943}. Its position is consistent with the unidentified hard X-ray source \textit{IGR J19443+2117}. The source is detected at $7.9\sigma$ in ~$25$ hours of observation live-time: its flux corresponds to $\sim0.02$ Crab units, with a photon index of $3.1\pm0.3_{stat}$ between $470$ GeV and $6$ TeV. \\
The study of the spectral energy distribution of this source, including radio, infrared and X-ray data, suggests an extra-galactic origin of the emission. A gamma-ray binary hypothesis is disfavored mainly due to the lack of a plausible massive stellar counterpart in optical/infrared and the absence of orbital variability. A pulsar wind nebula origin is disfavored by the very soft TeV spectrum and the absence of extended X-ray counterparts. On the other hand, the overall spectral energy distribution is consistent with an extreme high-energy-peaked blazar, with a synchrotron peak energy $>1$ keV. The lower limit on the redshift, based on the expected flux from the host galaxy, is 
$z>0.14$.

\section{Summary and conclusions}

A summary of the \hess\ extra-galactic sky was presented, with a particular emphasis given on the latest TeV-emitting AGNs detected by \hess\ .\\
The increasing number of detections claimed by \hess\ in the last years is mainly due to the development of new high-performance analysis techniques. In the near future, the study of the VHE sky will be significantly improved with the construction of a fifth 24m-diameter telescope : in this new configuration (\hess\ \textit{II}), the increasing of sensitivity and the lowering of the energy threshold (down to $30$ GeV), will permit the detection of fainter and more distant objects, populating the known VHE extra-galactic sky. \hess\ \textit{II} is currently under construction and should be fully operational by the end of 2012.

\section{Acknowledgements}
The support of the Namibian authorities and of the University of Namibia in facilitating the
construction and operation of \hess\ is gratefully acknowledged, as is the support by the German
Ministry for Education and Research (BMBF), the Max Planck Society, the French Ministry
for Research, the CNRS-IN2P3 and the Astroparticle Interdisciplinary Programme of the CNRS,
the U.K. Science and Technology Facilities Council (STFC), the IPNP of the Charles University,
the Polish Ministry of Science and Higher Education, the South African Department of Science
and Technology and National Research Foundation, and by the University of Namibia. We appreciate
the excellent work of the technical support staff in Berlin, Durham, Hamburg, Heidelberg,
Palaiseau, Paris, Saclay, and in Namibia in the construction and operation of the equipment.



\clearpage

\end{document}